# Large Magnetic-Field-Induced Strain at the Magnetic Order Transition in Triangular Antiferromagnet AgCrS$_2$

Tomoya Kanematsu[1], Yoshihiko Okamoto,[1,a)] and Koshi Takenaka[1]

[1]*Department of Applied Physics, Nagoya University, Nagoya 464-8603, Japan*

Strain induced by a magnetic field is a common phenomenon for ferromagnets, but few antiferromagnets show large strain induced by a magnetic field. On the basis of linear strain measurements of sintered samples of triangular antiferromagnet $A$CrS$_2$ ($A$ = Cu, Ag, and Au) in magnetic fields up to 9 T, the AgCrS$_2$ sample was found to show a large strain, yielding a large volume change over 700 ppm, which is one of the largest volume changes measured to date for an antiferromagnet. This large strain appeared only at the Néel temperature of 42 K and was not restored to its initial state when the applied magnetic field was decreased to zero; however, it was initialized by cooling the sample to far below the Néel temperature. These results suggest that the coexistence of magnetically ordered and paramagnetic phases at the first-order phase transition plays an important role. AuCrS$_2$ showed a magnetic-field-induced strain with similar features, although it was smaller than that in AgCrS$_2$.

There is a long history of studying the strain induced by applying a magnetic field to ferromagnetic materials. The most classical example is the magnetostriction of elemental metals, such as iron and nickel, which show elastic strains of several to tens of ppm.[1] Subsequent studies have discovered giant magnetostrictive materials such as Terfenol-D and Galfenol,[1-5] which have been put into practical use as ultrasonic transducers and actuators. Ferromagnetic shape-memory alloys, such as Ni$_2$MnGa and Fe$_3$Pt, are also known to show huge strains induced by magnetic fields.[6-9] Their strains are thermoelastic, i.e., they remain even if the applied magnetic field is decreased to zero, but they can be initialized by increasing the temperature to the austenite phase. The strains in the aforementioned ferromagnetic materials are basically directional magnetostrictions, where volume changes are small. In contrast, invar alloys have shown large magnetic-field-induced strains accompanied by a large volume changes, which is known as a forced volume magnetostriction.[10,11]

Unlike ferromagnetic materials, very few antiferromagnets show large magnetic-field-induced strains. There are some antiferromagnetic metals that show a large volume change associated with the antiferromagnetic order, such as YMn$_2$ and Mn$_3$GaN, but their response to a magnetic field is small.[12-15] ZnCr$_2$Se$_4$ and LiInCr$_4$S$_8$ are rare examples that show large magnetic-field-induced strains. These materials are spin-3/2 antiferromagnets, and show an antiferromagnetic order accompanied by a large volume change at Néel temperature of $T_N$ = 21 K and 24 K, respectively.[16-18] Sintered samples of these materials showed elastic strains of up to 660 ppm and 200 ppm at low temperatures by applying a magnetic field from 0 T to 7 T, respectively. The strain in LiInCr$_4$S$_8$ is isotropic, giving rise to a large volume change. It was pointed out that the large shift of $T_N$ by applying a magnetic field and the presence of strong ferromagnetic interaction play important roles in the large strains in these Cr spinel compounds.[16,18] On the other hand, antiferromagnets with spontaneous magnetization, such as canted antiferromagnets, often show large magnetic-field-induced strains.[19,20]

In this paper, we focus on magnetic-field-induced strains in $A$CrS$_2$ ($A$ = Cu, Ag, and Au). $A$CrS$_2$ has a layered crystal structure, where CrS$_2$ layers comprising edge-shared CrS$_6$ octahedra are separated by $A^+$ atoms. In a CrS$_2$ layer, Cr$^{3+}$ ions with $S$ = 3/2 spin form a triangular lattice. The crystal structures of $A$ = Cu and Ag have a noncentrosymmetric $R3m$ space group at room temperature, while that of AuCrS$_2$ is $R\bar{3}m$ with an inversion center.[21,22] This difference depends on whether the $A^+$ atom is tetrahedrally or linearly coordinated by sulfur atoms. In these three compounds, antiferromagnetic and ferromagnetic interactions coexist, but the antiferromagnetic interaction is dominant, as indicated by their negative Weiss temperatures, $\theta_W$.[21,22] $A$CrS$_2$ has been reported to exhibit an antiferromagnetic order without spontaneous magnetization at $T_N$ = 37−39 K, 41−42 K, and 47 K for $A$ = Cu, Ag, and Au, respectively.[21,23-27] This order was accompanied by a structural transition with large changes in their lattice constants, resulting in monoclinic below $T_N$.[24-26,28] No dilatometric measurements have been made on $A$CrS$_2$. The neutron scattering data of $A$ = Ag and Au have indicated that their spin structures below $T_N$ have an uudd configuration in a triangular-lattice plane.[24,26] In contrast, CuCrS$_2$ was pointed out to have a helical spin structure.[29] Here, we report

―――――――――――――――――――
a) Electronic mail: yokamoto@nuap.nagoya-u.ac.jp

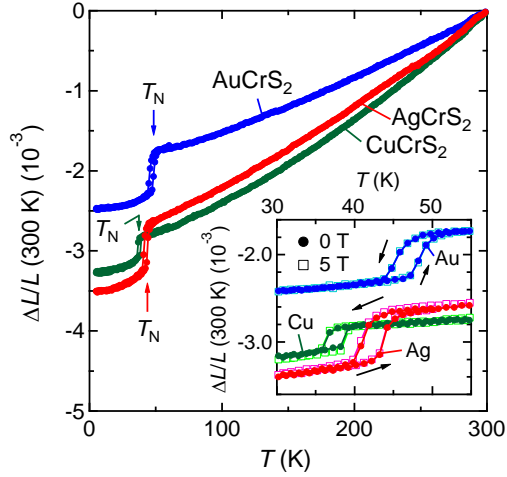

FIG. 1. Linear thermal expansion of the $A$CrS$_2$ ($A$ = Cu, Ag, and Au) sintered samples normalized to 300 K. The inset shows the data around $T_N$. The filled and open symbols show the data measured at zero magnetic field and 5 T, respectively. For each sample, the linear thermal expansion in the cooling process was first measured, and then that in the heating process was measured.

that sintered samples of AgCrS$_2$ show large magnetic-field-induced strains only at around $T_N$, which results in a volume increase exceeding 700 ppm at 9 T. This strain was not completely restored by decreasing the magnetic field to zero, but it was initialized by cooling the sample well below $T_N$, indicating that the strain includes both elastic and thermoelastic contributions. The AuCrS$_2$ sample showed a magnetic-field-induced strain with similar features, although it was smaller than that in AgCrS$_2$. In contrast, CuCrS$_2$ did not show any significant strain in the applied magnetic fields.

Sintered samples of $A$CrS$_2$ ($A$ = Cu, Ag, and Au) were prepared by a solid-state reaction method (See Supplementary Note 1). Linear thermal expansion and linear strain in magnetic fields of the sintered samples were measured using a strain gauge (KFLB, 120 W, Kyowa Electronic Instruments Co.) with a Cu reference.[30] The uncertainty in the linear thermal expansion and linear strain data is at most a few tens of ppm. In linear thermal expansion measurements, there is a relation of $\Delta V/V = 3(\Delta L/L)$ between the volume change $\Delta V/V$ and linear strain $\Delta L/L$. On the other hand, a volume change in a magnetic field is represented as $\Delta V/V = 2(\Delta L/L)_\perp + (\Delta L/L)_{//}$, where $(\Delta L/L)_\perp$ and $(\Delta L/L)_{//}$ are the linear strains measured perpendicular and parallel to the magnetic field, respectively. Linear strain in the presence of a magnetic field, thermal expansion, and heat capacity were measured using a Physical Property Measurement System (Quantum Design).

Figure 1 shows the linear thermal expansion, $\Delta L/L$, of the sintered samples of $A$CrS$_2$ ($A$ = Cu, Ag, and Au). In all samples, $\Delta L/L$ sharply decreased at $T_N$ with decreasing temperature, which was due to the structural change associated with the antiferromagnetic order. As shown in the inset of Fig. 1, there was a temperature hysteresis of several K in each sample, reflecting the first-order phase transition. The data measured at zero magnetic field and 5 T were almost identical, indicating that the thermal expansion was insensitive to the magnetic field. The volume changes at this anomaly were $\Delta V/V = 3(\Delta L/L) = 800$ ppm, 1800 ppm, and 1500 ppm for $A$ = Cu, Ag, and Au, respectively. The $\Delta V/V$ values of AgCrS$_2$ and AuCrS$_2$ were smaller than the crystallographic volume changes of 3400 ppm and 1900 ppm, respectively, which were estimated using the lattice constants above and below $T_N$.[26,28] This difference may suggest the presence of microstructural effects caused by the crystal grains with anisotropic thermal expansion and the pores in ceramics.[31-33] At $T > T_N$, the AgCrS$_2$ and CuCrS$_2$ samples showed similar $\Delta L/L$, but AuCrS$_2$ showed considerably smaller thermal expansion, which might be due to the different crystal structures; the $A$ atom is tetrahedrally coordinated by sulfur atoms in CuCrS$_2$ and AgCrS$_2$, while linearly coordinated in AuCrS$_2$, suggesting that the covalency in AuCrS$_2$ is stronger than those in the other two compounds.

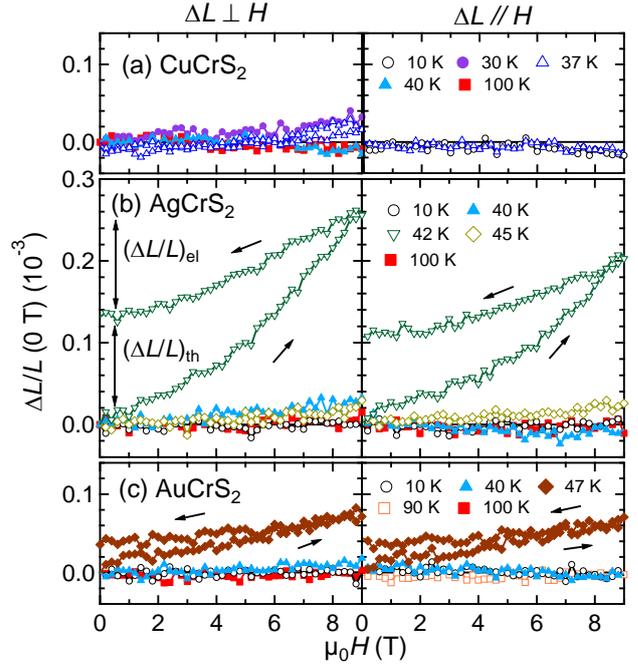

FIG. 2. Linear strains of the sintered samples of CuCrS$_2$ (a), AgCrS$_2$ (b), and AuCrS$_2$ (c) measured in magnetic fields. For each sample, the 10 K data were first measured after cooling the sample to 10 K at zero magnetic field, and then other data were measured in order of increasing temperature without kept at other temperatures. All data were normalized to the zero-field data before applying the magnetic field at each temperature. The left and right panels show the transverse and longitudinal magnetostrictions, where linear strains were measured perpendicular to and along the magnetic fields, respectively.

Figure 2 shows the linear strains of the $A$CrS$_2$ sintered samples measured in magnetic fields. As shown in Fig. 2(a),

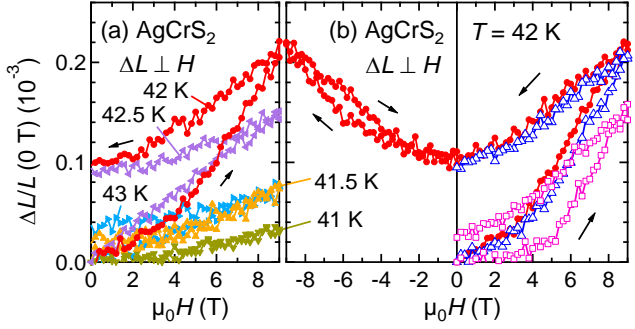

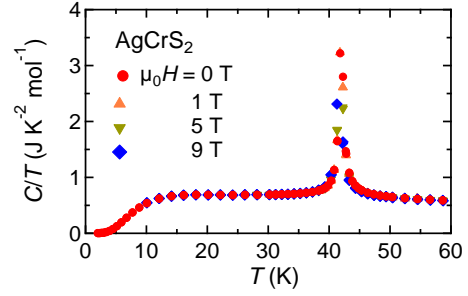

FIG. 3. Magnetic-field dependences of the linear strain of the sintered sample of AgCrS$_2$ measured perpendicular to the magnetic field. (a) The linear strain between 41 K and 43 K at intervals of 0.5 K. For each temperature, the linear strain was measured with increasing and decreasing magnetic fields between 0 T and 9 T, after the sample was warmed from 10 K at zero magnetic field. (b) The linear strain at 42 K. Filled circles show the linear strain data measured in the process of warming the sample from low temperature to 42 K in the zero magnetic field, increasing the magnetic field from 0 T to 9 T, decreasing it from 9 T to −9 T, and then increasing it again to 0 T. This sample was then warmed to 100 K, cooled to 42 K, and then the linear strain between 0 T and 9 T was measured (open squares). Finally, this sample was cooled to 10 K, warmed to 42 K again, and then the linear strain between 0 T and 9 T was measured (open triangles). The data in each dataset were normalized to the zero-field data before applying the magnetic field.

the CuCrS$_2$ sample showed small $\Delta L/L$ of up to 30 ppm at 9 T. In contrast, as shown in Fig. 2(b), a large strain was observed in the AgCrS$_2$ sample at 42 K. As seen in Fig. 3(a), the largest strain was realized at 42 K. The linear strain from 0 T to 9 T was $(\Delta L/L)_\perp = 260$ ppm and $(\Delta L/L)_{//} = 210$ ppm, yielding a large volume increase of $\Delta V/V = 730$ ppm. This $\Delta V/V$ is comparable to $\Delta V/V = 780$ ppm in LiInCr$_4$S$_8$, which is one of the largest volume changes yet seen for an antiferromagnet.[18] The magnetic-field-induced strain of AgCrS$_2$ at 42 K was not completely restored when the magnetic field was decreased to zero. In other words, the strain contained elastic $(\Delta L/L)_{el}$ and irreversible $(\Delta L/L)_{th}$ contributions, as indicated in the left panel of Fig. 2(b). As discussed later, $(\Delta L/L)_{th}$ is thermoelastic. At 41 K and 45 K, $\Delta L/L$ at 9 T was at most 30 ppm, indicating that a large strain appeared only in a narrow temperature range near $T_N$. As shown in Fig. 2(c), the AuCrS$_2$ sample showed a moderately large $\Delta L/L$ at $T_N = 47$ K, yielding $\Delta V/V = 210$ ppm at 9 T. This strain was irreversible, as in the case of AgCrS$_2$.

The elastic strain $(\Delta L/L)_{el}$ of AgCrS$_2$ at 42 K repeatedly appeared regardless of the temperature history, as shown in Fig. 3(b). The volume change due to $(\Delta L/L)_{el}$ between 0 T and 9 T was approximately 360 ppm. In contrast, $(\Delta L/L)_{th}$ strongly depended on the temperature history. When the sample was kept at 42 K, $(\Delta L/L)_{th}$ appeared only in the initial process

FIG. 4. Temperature dependence of heat capacity divided by temperature of the sintered sample of AgCrS$_2$ measured under various magnetic fields.

[filled circles in Fig. 3(b)]. However, after the sample was cooled to 10 K and then warmed to 42 K again, $(\Delta L/L)_{th}$ appeared in the same way as in the initial process (open tringles). In contrast, considerably smaller $(\Delta L/L)_{th}$ was observed after the sample was cooled from 100 K to 42 K (open squares).

However, the magnetic-field-induced strain in AgCrS$_2$ has different features from those in ZnCr$_2$Se$_4$ and LiInCr$_4$S$_8$. The strains in ZnCr$_2$Se$_4$ and LiInCr$_4$S$_8$ are mainly due to an antiferromagnetic-paramagnetic phase transition induced by an applied magnetic field. The $T_N$ of ZnCr$_2$Se$_4$ was 21 K in a zero magnetic field. By applying a magnetic field of 7 T, the $T_N$ was decreased below 5 K. Correspondingly, a large magnetic-field-induced strain appeared in a wide temperature range below $T_N$, caused by crossing the magnetic phase boundary by applying a magnetic field.[16] The $T_N$ of LiInCr$_4$S$_8$ decreased by several K from 24 K at 5 T, and a large strain was induced by applying a magnetic field at 23 K just below $T_N$, which is in contrast to the case of the $2p$ analogue LiInCr$_4$O$_8$.[18,34] These strains were reversible with respect to any change of the magnetic field, reflecting the phase transition between thermal equilibrium states. In contrast, the $T_N$ of AgCrS$_2$ was insensitive to magnetic fields up to 9 T, as seen in the heat capacity data shown in Fig. 4 and the $\Delta L/L$ data shown in the inset of Fig. 1. These data suggest that the phase transition in AgCrS$_2$ could not be induced by a magnetic field of 9 T, and therefore the large magnetic-field-induced strain in AgCrS$_2$ occurred by a different mechanism from the cases of ZnCr$_2$Se$_4$ and LiInCr$_4$S$_8$.

The large magnetic-field-induced strain in AgCrS$_2$ appeared only at 42 K and 42.5 K, as shown in Fig. 3(a), which was in the temperature hysteresis of $\Delta L/L$ shown in the inset of Fig. 1. Further, this strain included not only elastic $(\Delta L/L)_{el}$, but also irreversible $(\Delta L/L)_{th}$. These experimental results clearly showed that the coexistence of the magnetically ordered phase and the paramagnetic phase at the first-order phase transition played an essential role in the large magnetic-field-induced strain in AgCrS$_2$. As seen in Fig. 1, $\Delta L/L$ discontinuously changed at $T_N$, and the volume of the paramagnetic phase was approximately 1800 ppm larger than that of the magnetically ordered phase. After the sample was warmed from low temperature to 42 K, most of the sample was expected to be in the magnetically ordered phase. When a magnetic field was then applied, a part of the sample was changed to the paramagnetic phase, resulting in a large volume increase.

Since this magnetic-field-induced change was not a phase transition between thermal equilibrium states, it could be irreversible in contrast to those of $ZnCr_2Se_4$ and $LiInCr_4S_8$. The presence of $(\Delta L/L)_{th}$ reflects it. This $(\Delta L/L)_{th}$ was initialized by keeping the sample at a sufficiently low temperature, which changed the whole sample into a magnetically ordered phase. Such an initialization process is similar to the case of ferromagnetic shape-memory alloys,[6,7] so $(\Delta L/L)_{th}$ is thermoelastic. This feature also appeared in the magnetization process of $AgCrS_2$ at 42 K, as described in Supplementary Note 2. In addition, when the sample was warmed from low temperature to 42 K, and then a magnetic field was applied and maintained at 9 T, $\Delta L/L$ continued to increase gradually (See Supplementary Fig. 3). This behavior is similar to that seen for $Mn_3CuN$,[35] indicating that this phenomenon is not purely electronic but might be due to some mechanical effects, such as the stresses associated with the formation and/or movement of microdomains and the effects of grain boundaries. This result supports the notion that the changes of volume fraction of the two phases in the two-phase coexistence state appeared as a large strain in $AgCrS_2$. However, $(\Delta L/L)_{el}$ can also be understood by this change, but it might not be the only mechanism for $(\Delta L/L)_{el}$. The forced volume magnetostriction of each phase itself could contribute to $(\Delta L/L)_{el}$, although it is not expected to be the main mechanism because the large strains appeared only at $T_N$.

$AgCrS_2$ is a rare antiferromagnet that has exhibited a large and thermoelastic strain induced by applying a magnetic field. The strain in $AuCrS_2$ had similar thermoelastic features as seen for $AgCrS_2$, but was much smaller. $CuCrS_2$ did not show a significant strain under application of a magnetic field. These three compounds have similar $T_N$ values and show similar discontinuous volume changes at $T_N$, as shown in Fig. 1. An important difference in the magnetic properties of $ACrS_2$ is the magnitude of the magnetic interactions. The $|\theta_W|$ = 55 K of $AgCrS_2$ was about half as large as those of $CuCrS_2$ and $AuCrS_2$, suggesting that the ferromagnetic interaction in $AgCrS_2$ is relatively stronger than those in the other two compounds.[21,25,26] The presence of a considerably strong ferromagnetic interaction in $AgCrS_2$ was also evidenced by first-principles calculations.[36] The average magnetic interaction in $AgCrS_2$ was antiferromagnetic, as indicated by the negative $\theta_W$. However, due to the geometrical frustration of the triangular lattice, the influence of the ferromagnetic interaction on the physical properties was expected to be more remarkable, which might appear as the large magnetic-field-induced strain in $AgCrS_2$. The different magnetic structures below $T_N$ might also have had an effect. The magnetic structure in $AgCrS_2$ and $AuCrS_2$ was reported to be an uudd configuration in a triangular plane, while that in $CuCrS_2$ was reported to be spiral.[24,26,29] We hope that future research will elucidate the reason why a large strain was realized only in $AgCrS_2$, which may lead to the discovery of an antiferromagnetic material that exhibits a large magnetic-field-induced strain over a wider temperature range, which could be an actuator material possessing no spontaneous magnetization.

In summary, we found that sintered samples of $AgCrS_2$ showed a large magnetic-field-induced strain at $T_N$ = 42 K, which was accompanied by a large volume increase over 700 ppm by applying a magnetic field from 0 T to 9 T. This strain is unique in that it is not only a very large strain for an antiferromagnet, but it also has a thermoelastic nature; it was not completely restored to its initial state when the magnetic field was decreased to zero, but it was initialized by cooling the sample well below $T_N$. The $AuCrS_2$ sample showed a magnetic-field-induced strain with similar features, although the strain was considerably smaller than that in $AgCrS_2$. $CuCrS_2$ did not show significant $\Delta L/L$ in magnetic fields. The coexistence of the magnetically ordered and paramagnetic phases at the first-order phase transition and the strong competition between antiferromagnetic and ferromagnetic interactions on the geometrically frustrated lattice were expected to play essential roles in the large magnetic-field-induced strain in $ACrS_2$.

## ACKNOWLEDGMENTS

The authors are grateful to K. Eto and Y. Kubota for their help with experiments and D. Hirai and Z. Hiroi for their support in heat capacity and magnetization measurements. The work was partly carried out under the Visiting Researcher Program of the Institute for Solid State Physics, the University of Tokyo and JSPS KAKENHI (Grant Numbers: 19H05823, 20H00346 and 20H02603).

## REFERENCES

[1] A. E. Clark, *Handbook of Ferromagnetic Materials* (North-Holland Publishing Company, Amsterdam, 1980), Vol. 1, pp. 531-589.
[2] A. E. Clark, J. P. Teter, and O. D. McMasters, J. Appl. Phys. **63**, 3910 (1988).
[3] A. E. Clark, M. Wun-Fogle, J. B. Restorff, and T. A. Lograsso, Mater. Trans. **43**, 881 (2002).
[4] A. E. Clark, K. B. Hathaway, M. Wun-Fogle, J. B. Restorff, T. A. Lograsso, V. M. Keppens, G. Petculescu, and R. A. Taylor, J. Appl. Phys. **93**, 8621 (2003).
[5] R. A. Kellogg, A. Flatau, A. E. Clark, M. Wun-Fogle, T. Lograsso, J. Intell. Mater. Syst. Struct. **16**, 471 (2005).
[6] K. Ullakko, J. K. Huang, C. Kantner, R. C. O'Handley, and V. V. Kokorin, Appl. Phys. Lett. **69**, 1966 (1996).
[7] S. J. Murray, M. Marioni, S. M. Allen, R. C. O'Handley, and T. A.

[8]T. Sakamoto, T. Fukuda, T. Kakeshita, T. Takeuchi, and K. Kishio, J. Appl. Phys. **93**, 8647 (2003).

[9]T. Kakeshita, T. Takeuchi, T. Fukuda, M. Tsujiguchi, T. Saburi, R. Oshima and S. Muto, Appl. Phys. Lett. **77**, 1502 (2000).

[10]H. Nagaoka and K. Honda, Phil. Mag. **4**, 45 (1902).

[11]M. Matsumoto, T. Kaneko, and H. Fujimori, J. Phys. Soc. Jpn. **26**, 1083 (1969).

[12]M. Shiga, H. Wada, and Y. Nakamura, J. Mag. Mag. Mater. **31-34**, 119 (1983).

[13]Y. Nakamura, J. Magn. Magn. Mater. **31-34**, 829 (1983).

[14]Ph. l'Heritier, D. Boursier, R. Fruchart, and D. Fruchart, Mat. Res. Bull. **14**, 1203 (1979).

[15]K. Takenaka and H. Takagi, Appl. Phys. Lett. **87**, 261902 (2005).

[16]J. Hemberger, H.-A. Krug von Nidda, V. Tsurkan, and A. Loidl, Phys. Rev. Lett. **98**, 147203 (2007).

[17]Y. Okamoto, M. Mori, N. Katayama, A. Miyake, M. Tokunaga, A. Matsuo, K. Kindo, and K. Takenaka, J. Phys. Soc. Jpn. **87**, 034709 (2018).

[18]T. Kanematsu, M. Mori, Y. Okamoto, T. Yajima, and K. Takenaka, J. Phys. Soc. Jpn. **89**, 073708 (2020).

[19]J. Baier, D. Meier, K. Berggold, J. Hemberger, A. Balbashov, J. A. Mydosh, and T. Lorenz, Phys. Rev. B **73**, 100402 (2006).

[20]R. Mahendiran, M. R. Ibarra, C. Marquina, B. Garcia-Landa, L. Morellon, A. Maignan, B. Raveau, A. Arulraj, and C. N. R. Rao, Appl. Phys. Lett. **82**, 242 (2003).

[21]P. F. Bongers, C. F. Van Bruggen, J. Koopstra, W. P. F. A. M. Omloo, G. A. Wiegers, and F. Jellinek, J. Phys. Chem. Solids **29**, 977 (1968).

[22]H. Fukuoka, S. Sakashita, and S. Yamanaka, J. Solid State Chem. **148**, 487 (1999).

[23]A. Karmakar, K. Dey, S. Chatterjee, S. Majumdar, and S. Giri, Appl. Phys. Lett. **104**, 052906 (2004).

[24]F. Damay, C. Martin, V. Hardy, G. André, S. Petit, and A. Maignan, Phys. Rev. B **83**, 184413 (2011).

[25]J. C. E. Rasch, M. Boehm, C. Ritter, H. Mutka, J. Schefer, L. Keller, G. M. Abramova, A. Cervellino, and J. F. Löffler, Phys. Rev. B **80**, 104431 (2009).

[26]S. J. E. Carlsson, G. Rousse, I. Yamada, H. Kuriki, R. Takahashi, F. Lévy-Bertrand, G. Giriat, and A. Gauzzi, Phys. Rev. B **84**, 094455 (2011).

[27]K. Singh, A. Maignan, C. Martin, and Ch. Simon, Chem. Mater. **21**, 5007 (2009).

[28]K. R. S. Preethi Meher, C. Martin, V. Caignaert, F. Damay, and A. Maignan, Chem. Mater. **26**, 830 (2014).

[29]M. Winterberger and Y. Allain, Solid State Commun. **64**, 1343 (1987).

[30]K. Takenaka, M. Ichigo, T. Hamada, A. Ozawa, T. Shibayama, T. Inagaki, and K. Asano, Sci. Technol. Adv. Mater. **15**, 015009 (2014).

[31]F. H. Gillery and E. A. Bush, J. Am. Ceram. Soc. **42**, 175 (1959).

[32]K. Takenaka, Y. Okamoto, T. Shinoda, N. Katayama, and Y. Sakai, Nat. Commun. **8**, 14102 (2017).

[33]N. Katayama, K. Otsuka, M. Mitamura, Y. Yokoyama, Y. Okamoto, and K. Takenaka, Appl. Phys. Lett. **113**, 181902 (2018).

[34]Y. Okamoto, G. J. Nilsen, J. P. Attfield, and Z. Hiroi, Phys. Rev. Lett. **110**, 097203 (2013).

[35]K. Asano, K. Koyama, and K. Takenaka, Appl. Phys. Lett. **92**, 161909 (2008).

[36]A. V. Ushakov, D. A. Kukusta, A. N. Yaresko, and D. I. Khomskii, Phys. Rev. B **87**, 014418 (2013).

**Supplementary Note 1. Sample preparation**

Sintered samples of $A$CrS$_2$ ($A$ = Cu, Ag, and Au) were prepared by a solid-state reaction method. A stoichiometric amount of elemental powders were mixed, pressed, and then sealed in an evacuated quartz tube. The tube was heated to and kept at 673 K for 24 h and then at 1273 K for 48 h, 1173 K for 12 h, and 1073 K for 48 h for $A$ = Cu, Ag, and Au, respectively. Thereafter, the tube was furnace cooled to room temperature. Sample characterization was performed by powder X-ray diffraction (XRD) with Cu Kα radiation at room temperature using a MiniFlex diffractometer (RIGAKU). Supplementary Fig. 1 shows powder XRD patterns of the sintered samples of $A$CrS$_2$ ($A$ = Cu, Ag, and Au) measured at room temperature. For each pattern, all the peaks, except for the small peaks from the impurity phase of gold in AuCrS$_2$, were indexed on the basis of trigonal unit cells with lattice constants $a$ = 3.4704(15) Å, 3.4987(6) Å, and 3.502(3) Å and $c$ = 18.632 (9) Å, 20.549 (5) Å, and 21.531(10) Å for $A$ = Cu, Ag, and Au, respectively. These lattice constants were almost the same as those reported in previous studies,[22,24,25] indicating that the CuCrS$_2$, and the AgCrS$_2$ sintered samples are a single phase of $A$CrS$_2$ and AuCrS$_2$ sintered sample was an almost single phase. The facts that the prepared samples in this study contained no or few amounts of impurity phases and no elemental sulfur remained in the quartz tube after the reaction suggest that the chemical compositions of the sintered samples are stoichiometric.

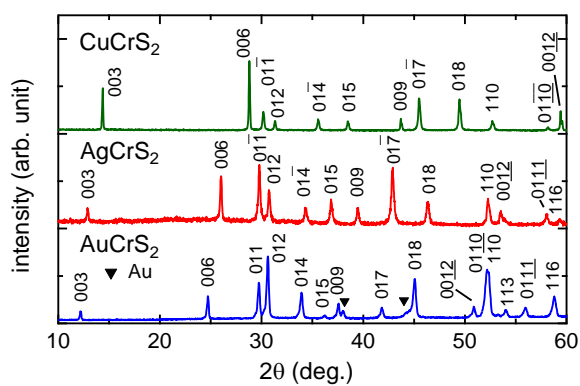

Supplementary Figure 1. Powder XRD patterns of the sintered samples of $A$CrS$_2$ ($A$ = Cu, Ag, and Au) measured at room temperature. Filled triangles indicate the diffraction peaks of an impurity phase of gold. Peak indices are given using trigonal unit cells with lattice constants of $a$ = 3.4704(15) Å, 3.4987(6) Å, and 3.502(3) Å and $c$ = 18.632(9) Å, 20.549(5) Å, and 21.531(10) Å for $A$ = Cu, Ag, and Au, respectively.

## Supplementary Note 2. Magnetization process of a sintered sample of AgCrS$_2$

Supplementary Fig. 2 shows the magnetization process of the AgCrS$_2$ sintered sample. The data for both increasing and decreasing magnetic fields are shown. Magnetization measurements were performed using an MPMS-3 (Quantum Design). At 42 K, the first magnetization curve measured after the sample was warmed from 10 K to 42 K (filled circles) shows different values from those of the second curve measured just after the measurement of the first curve where the sample was maintained at 42 K (open triangles).

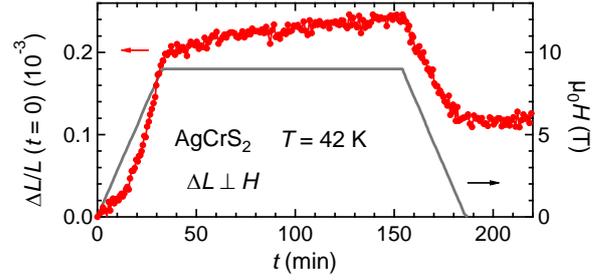

Supplementary Figure 3. Time dependence of the linear strain of an AgCrS$_2$ sintered sample measured perpendicular to the magnetic field. The time variation of the magnetic field is shown on the right axis.

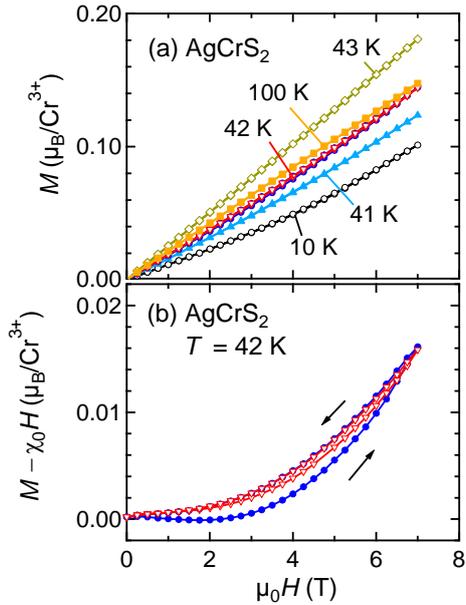

Supplementary Figure 2. Magnetization processes of the AgCrS$_2$ sintered sample. (a) Magnetization processes measured at 10, 41, 42, 43, and 100 K. (b) Magnetization processes at 42 K after subtracting the linear contribution of the magnetic field ($\chi_0 = 1.024 \times 10^{-2}$ cm$^3$ mol$^{-1}$). Filled circles show the first magnetization process measured after warming the sample from 10 K to 42 K in a zero magnetic field, while open triangles show the process measured just after the first process where the sample was maintained at 42 K.

## Supplementary Note 3. Time dependence of the magnetic-field-induced strain in a sintered sample of AgCrS$_2$

Supplementary Fig. 3 shows the time dependence of the linear strain of the AgCrS$_2$ sintered sample measured perpendicular to the magnetic field. The sample was warmed from low temperature to 42 K in the zero magnetic field. Then, the magnetic field was increased to 9 T at a rate of 0.3 T min$^{-1}$ from $t = 0$, kept at 9 T for 120 min, decreased to 0 T at a rate of $-0.3$ T min$^{-1}$, and then kept at 0 T for 30 min. The sample showed a large strain of 200 ppm with increasing magnetic field from 0 T to 9 T, and $\Delta L/L$ gradually increased at 9 T over time. After that, the sample showed a strain of $-120$ ppm with decreasing a magnetic field from 9 T to 0 T.